\documentclass[emulateapj,apj]{emulateapj}

\slugcomment{Not to appear in Nonlearned J., 45.}

\shorttitle{Gravitational Fragmentation of Expanding Shells. II.}
\shortauthors{Iwasaki, Inutsuka, and Tsuribe}

\begin{document}


\title{GRAVITATIONAL FRAGMENTATION OF EXPANDING SHELLS. II.
THREE-DIMENSIONAL SIMULATIONS}


\author{Kazunari Iwasaki\altaffilmark{1}, Shu-ichiro 
Inutsuka\altaffilmark{1}, and 
Toru Tsuribe\altaffilmark{2}}

\altaffiltext{1}{Department of Physics, Nagoya University, Furo-cho, 
Chikusa-ku, Nagoya, Aichi, 464-8602, Japan; 
iwasaki@nagoya-u.jp, inutsuka@nagoya-u.jp}
\altaffiltext{2}{Department of Earth and Space Science, Osaka University, 
Machikaneyama-cho 1-1, Toyonaka, Osaka, 
560-0043, Japan; tsuribe@vega.ess.sci.osaka-u.ac.jp}




\begin{abstract}
We investigate the gravitational fragmentation of expanding shells driven 
by HII regions using 
the three-dimensional Lagrangian simulation codes based on the Riemann solver, called 
Godunov smoothed particle hydrodynamics.
The ambient gas is assumed to be uniform.
In order to attain high resolution to resolve the geometrically thin dense shell, 
we calculate not the whole but a part of the shell. 
We find that perturbations begin to grow  earlier than
the prediction of the linear analysis under the thin-shell approximation.
The wavenumber of the most unstable mode is larger than that 
in the thin-shell linear analysis.
The development of the gravitational instability is accompanied by 
the significant deformation of the contact discontinuity.
These results are consistent with a linear analysis presented by Iwasaki et al. 
that have taken into account the density profile across the thickness and 
approximate shock and contact discontinuity boundary conditions. 
We derive useful analytic formulae 
for the fragment scale and
the epoch when the gravitational instability begins to grow.
\end{abstract}


\section{Introduction}
Massive stars strongly disturb the interstellar medium (ISM) via
emission of ionizing photons, stellar winds, and  supernova explosions.
These processes produce overpressured hot bubbles
that expand into ambient interstellar clouds. 
At the same time, a shock wave sweeps up the ambient clouds into 
a dense shell.
The expanding shell strongly influences the dynamics of the ISM.
Especially, it is often supposed to 
trigger the formation of stars through the gravitational fragmentation \citep{EL77}.
\citet{HI05,HI06} investigated the dynamical expansion of the HII region, the 
photodissociation region, and the swept-up shell, solving 
the UV and far-UV radiative transfer and thermal and 
chemical processes in the one-dimensional (1D) hydrodynamics code.
They showed that the shell becomes cold and dense enough for the 
gravitational instability (GI) to take place owing to the reformation of molecules destructed 
by far-UV photons.
Numerous authors have discovered evidences for 
the star formation in shell-like molecular clouds around hot bubbles.
\citet{C06,C07} have compiled a catalogue of $\sim600$ shells from 
data in {\it Spitzer}-GLIMPSE survey.
Recently, \citet{Detal10} have investigated 102 samples identified as shells on the 
{\it Spitzer}-GLIMPSE images at 8.0$\mu$m.
They found that 86\% of the shells are associated with ionized gases, or HII regions. 
They obtained statistical properties of the triggered star formation, and 
suggested that more than a quarter of the shells may have triggered the formation of massive stars.
This suggests that the trigger star formation process may be important in the massive 
star formation.

To understand the triggered star formation,
it is important to investigate how and when the expanding shell fragments through the GI.
\citet{E94} and \citet{Wetal94} derived a simple dispersion relation taking into account 
the mass accretion and the dilution effect of perturbations owing to the expansion.
They assumed the thin-shell approximation, and 
neglect the boundary effect of the contact discontinuity (CD).
Recently, \citet[][hereafter Paper I]{IIT11}  investigated stability of expanding shells
taking into account asymmetric density profiles with imposing an approximate shock boundary 
condition on the leading surface and the CD boundary condition on the trailing surface.
They found that the dispersion relation and eigen-function strongly depends on 
the boundary conditions and the degree of asymmetry of the density profile.

Although many authors have studied fragmentation of shells 
by using linear analysis, it is still uncertain how and when shells fragment.
This is because the stability analysis of the evolving shell is 
difficult to perform without any approximations.
Therefore, time-dependent multi-dimensional simulations are crucial to investigate it.
To resolve propagating thin shells all the time, 
the Eulerian  adaptive mesh refinement (AMR) technique in the mesh 
code or the smoothed particle hydrodynamics (SPH) have been used.
However, since enormous meshes or SPH particles are required 
to resolve the thickness, 
researches based on numerical simulations are less advanced.
\citet{DBW07} investigated the gravitational fragmentation of the shell driven 
by the expansion of the HII region by using three-dimensional (3D)
SPH taking into account the radiative transfer of ionizing photons.
In their simulation, its thickness was not resolved sufficiently.
\citet{Detal09} have investigated fragmentation of shells with fixed 
mass that expand into a hot rarefied gas by using 
two different numerical schemes, the Eulerian AMR method 
(the {\sc flash} code) and the SPH method.
They considered shells confined by a constant pressure on both surfaces.
In the more realistic situation where shells are confined by 
the CD and the shock front (SF), 
multi-dimensional simulations including the 
self-gravity have not been performed yet.

In this paper, we perform the 3D simulation to investigate 
the fragmentation process of expanding shells through GI.
As numerical method, we adopt the 3D SPH method.
In order to attain high resolution enough to resolve the thickness, 
we calculate not the whole but a part of the shell. 
The outline of the paper is as follows:
in Section \ref{sec:simulations}, we describe our model and simulations in detail.
Brief review of Paper I is presented in Section \ref{sec:review}. 
The results of our simulations are shown in Section \ref{sec:sph result}.
In Sections \ref{sec:discuss} and \ref{sec:summary}, we present discussions and summaries of 
our paper.

\section{Simulations}\label{sec:simulations}
\subsection{Numerical Methods}\label{sec:num method}
We adopt the SPH method \citep[e.g.,][]{M92}.
Since SPH is Lagrangian particle method, it is one of the 
best method for problems where a wide low density region 
and geometrically thin shell coexist.
Instead of additional viscosity to handle shocks,
we use the ``Godunov SPH'' where the results of the Riemann problem 
are used to calculate the interaction between SPH particles \citep{I02}.
The tree method \citep{BH86} is used to calculate self-gravitational force.

\subsubsection{Treatment of Contact Discontinuity}\label{sec:sph}
In the standard SPH method \citep{M92}, 
the $i$-th particle has the mass of $m_i$, and its density 
is given by
\begin{equation}
        \rho_i = \sum_k m_k W({\bf x}_i-{\bf x}_k,\bar{h}_{ik}),
        \label{eoc sph}
\end{equation}
where $W({\bf x},h)$ is kernel function and we use the Gaussian kernel, $h$ is
the smoothing length, and $\bar{h}_{ik}$ is an average between 
$h_i$ and $h_k$. 
The equation of motion of the $i$-th particle is given by
\begin{equation}
        \frac{\Delta {\bf v}_i}{\Delta t}
        = - \sum_k m_k \left( \frac{P_i}{\rho_i^2} + \frac{P_k}{\rho_k^2} \right)
        \frac{\partial}{\partial {\bf x}_i}W({\bf x}_i - {\bf x}_k,
        \bar{h}_{ik}).
        \label{eom sph}
\end{equation}
The standard SPH has a shortcoming when calculating the CD
between a hot and a cold gas \citep{A07}. 
\citet{A07} reported that the artificial tension stabilizes 
the Kelvin-Helmholtz instability with a different densities in
the standard SPH.
In contrast, \citet{C10} have shown that 
Godunov SPH can describe the Kelvin-Helmholtz instability
reasonably well.

In addition,
\citet{TM05} and \cite{HA06} proposed a new method that treats the CD naturally in 
the context of a multi-phase fluid.
They modified Equations (\ref{eoc sph}) and 
(\ref{eom sph}) to treat the CD as follows:
\begin{equation}
        \rho_i = m_i \sum_k W({\bf x}_i-{\bf x}_k,\bar{h}_{ik}),
        \label{eoc sph modi}
\end{equation}
and 
\begin{eqnarray}
        \frac{\Delta {\bf v}_i}{\Delta t}
        &=& 
        - \frac{1}{m_i}\sum_k \left[ P_i\left(\frac{m_i}{\rho_i}\right)^2 
        + P_k\left(\frac{m_k}{\rho_k} \right)^2\right]\nonumber\\
        &&\hspace{3cm}\times\frac{\partial}{\partial {\bf x}_i}
        W({\bf x}_i - {\bf x}_k.
        \bar{h}_{ik}).
        \label{eom sph modi}
\end{eqnarray}
The term of $m_i/\rho_i$ represents the volume of $i$-th particle $\sim h_i^3$.
Therefore, if we determine particle mass of hot and cold gases so that its smoothing length 
distributes smoothly across the CD, the terms inside the bracket in Equation (\ref{eom sph modi}) 
is smooth across the CD, leading to much better behavior at the CD.
Their new method is very useful in problems where 
the position of the CD is known in advance as in our modeling.
Equation (\ref{eom sph modi}) satisfies the relation of 
$m_i\Delta {\bf v}_i = - m_k \Delta{\bf v}_k$, suggesting that the momentum conservation 
is guaranteed.
Furthermore, 
we apply the Godunov technique to Equation (\ref{eom sph modi}) as
\begin{eqnarray}
        \frac{\Delta {\bf v}_i}{\Delta t}
        &=& 
        - \frac{1}{m_i}\sum_k P^*_{ik}\left[ \left(\frac{m_i}{\rho_i}\right)^2 
        + \left(\frac{m_k}{\rho_k} \right)^2\right]\nonumber\\
        &&\hspace{3cm}\times\frac{\partial}{\partial {\bf x}_i}
        W({\bf x}_i - {\bf x}_k,
        \bar{h}_{ik}).
        \label{eom sph godunov}
\end{eqnarray}
where $P^*_{ik}$ is the result of the Riemann problem between 
the $i$-th and the $k$-th particles \citep{I02}.

\subsection{Models}\label{sec:models}
Massive stars emit ultraviolet photons ($h\nu>13.6$ eV) and produce  
HII regions around them.
Here, we consider a massive star that emits ionizing photons with the photon number luminosity 
$Q_\mathrm{UV}\:[\mathrm{s}^{-1}]$ 
into the ambient gas with the uniform density of $\rho_\mathrm{E}=mn_\mathrm{E}$, where 
$n_\mathrm{E}$ and $m$ are the number density and the mean mass of the ambient gas particle, 
respectively.

\begin{figure}[htpb]
        \begin{center}
            \includegraphics[width=8cm]{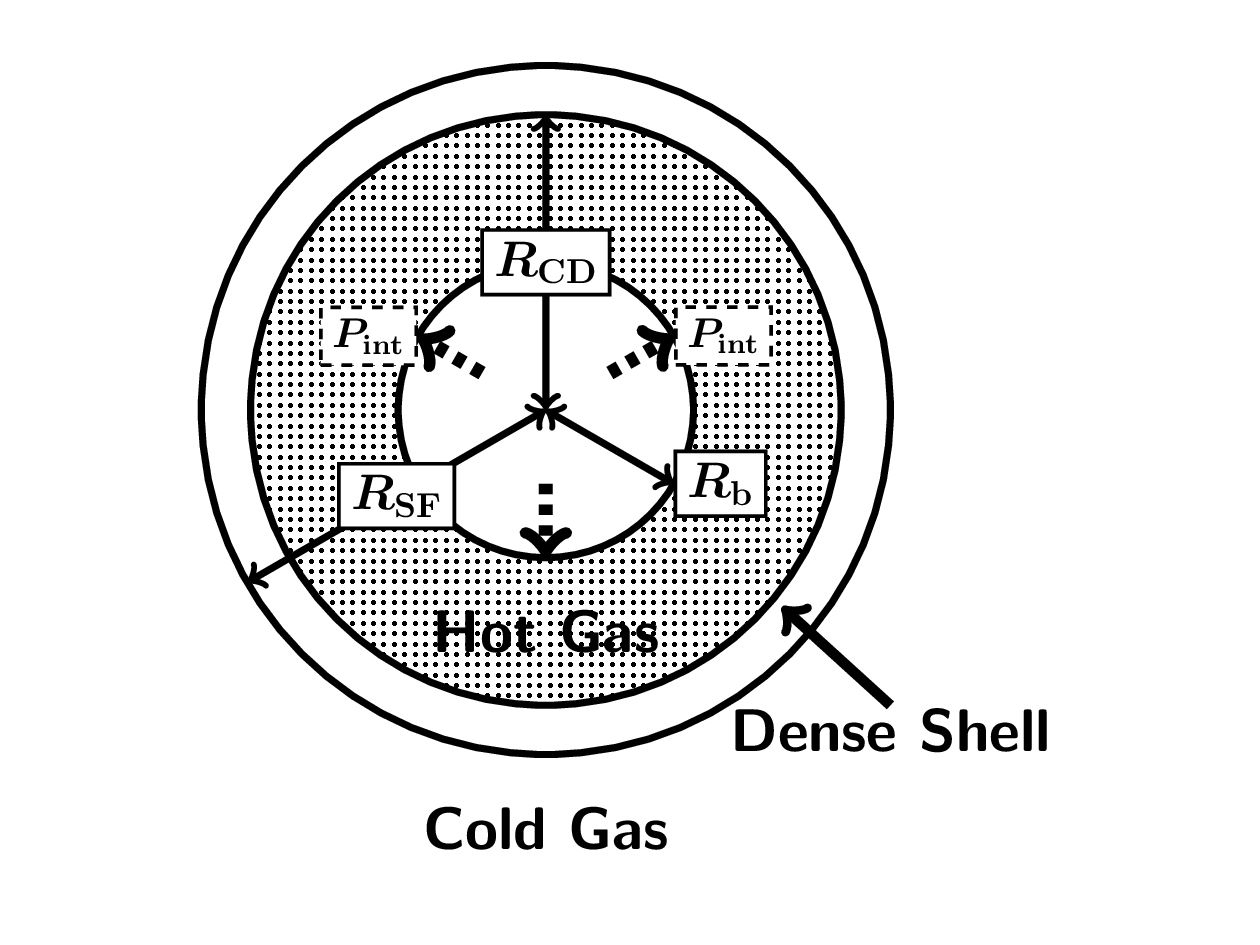}
        \end{center}
\caption{
         Schematic picture of our model. 
         The hot gas occupying the dotted region
         is embedded by the cold gas.
}
\label{setup}
\end{figure}

We construct as simple model as possible 
to concentrate on the physics of GI of expanding shells. 
Figure \ref{setup} illustrates the schematic picture of our model.
In this paper, we do not calculate the 
radiative transfer of ionizing photons but introduce hot and cold gases that
are assumed to evolve keeping their temperatures constant.
The hot gas is occupied in $r<R_\mathrm{CD}$ (the dotted region), and the thin shell is in 
$R_\mathrm{CD}<r<R_\mathrm{SF}$, where $R_\mathrm{CD}$ and 
$R_\mathrm{SF}$ are positions of the CD and the SF, respectively.
The pre-shock ambient cold gas in $r>R_\mathrm{SF}$ is assumed to be uniform.
In our model, the inner boundary is set at $r=R_\mathrm{b}$ inside the hot
bubble $(R_\mathrm{b}<R_\mathrm{CD})$.
In the HII region, the detailed balance between the recombination and the ionization 
is approximately established all the time.
Therefore, the interior pressure of the HII region is given by 
\begin{equation}
        P_\mathrm{int} = \rho_\mathrm{E} c_\mathrm{II}^2 \left( \frac{R_\mathrm{ST}}{R_\mathrm{CD}} \right)^{3/2},
    \label{Pext}
\end{equation}
where $R_\mathrm{ST}$ is the Str{\"o}mgren radius,
\begin{equation}
        R_\mathrm{ST} = \left( \frac{3Q_\mathrm{UV}}{4\pi \alpha_\mathrm{B} n_\mathrm{E}^2} \right)^{1/3}.
        \label{stremgren}
\end{equation}
We assume that the pressure of the HII region is spatially constant because of the high temperature.
The hot gas at $r=R_\mathrm{b}$ is pushed by the interior pressure $P_\mathrm{int}$.

\subsection{Calculation Domain}\label{sec:domain}
\begin{figure}[htpb]
        \begin{center}
            \includegraphics[width=8cm]{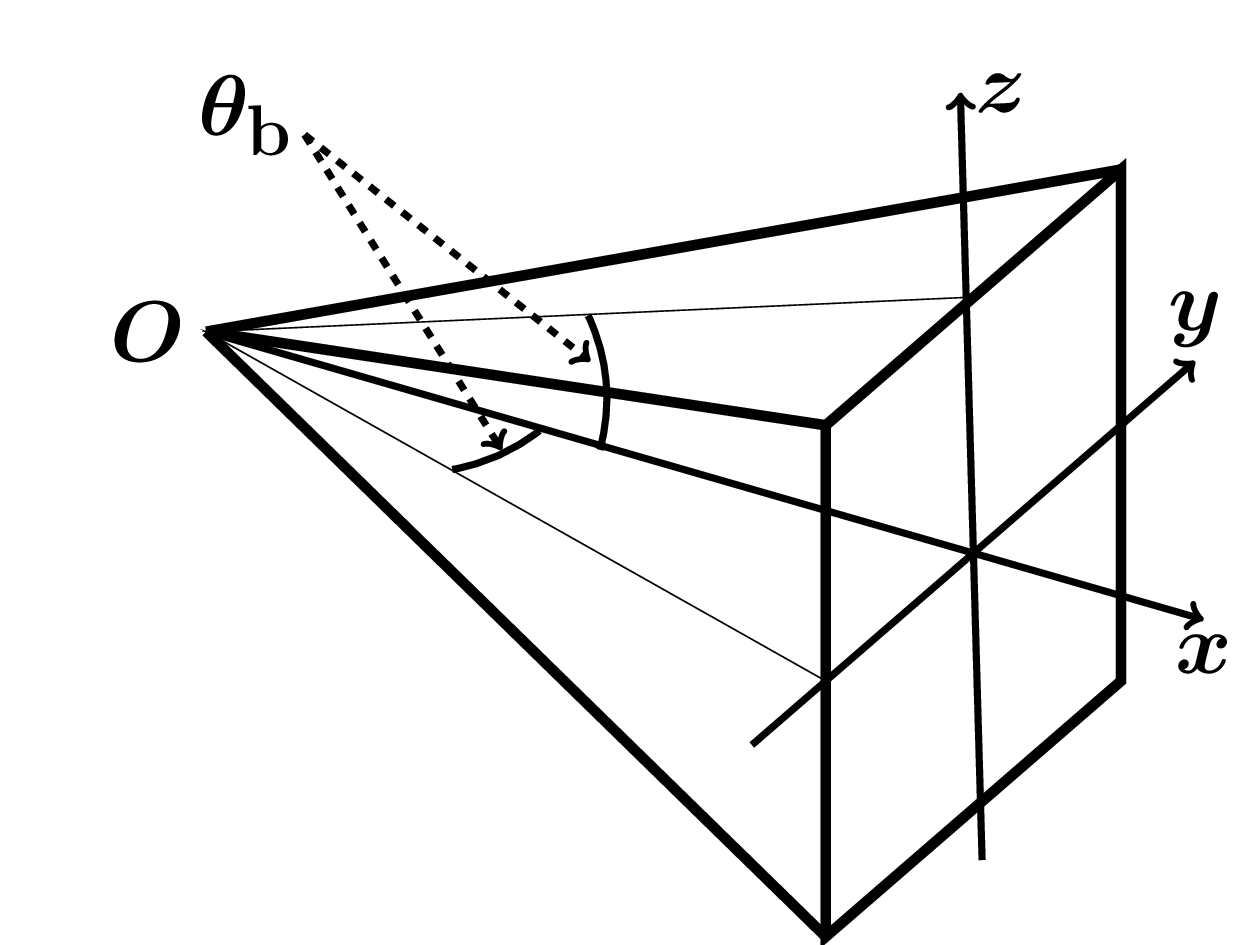}
        \end{center}
\caption{Schematic picture of the calculation domain.}
\label{fig:region}
\end{figure}

If we simulate the overall shell, the required number of 
SPH particles, $N_\mathrm{tot}$, is too large to calculate.
Therefore, to save $N_\mathrm{tot}$,
we calculate a part of the shell.
The schematic picture of the calculation domain is shown in 
Figure \ref{fig:region}, and 
is designated by 
$|\tan^{-1}(y/x)|\le\theta_\mathrm{b}$ and $|\tan^{-1}(z/x)|\le\theta_\mathrm{b}$.
Since the solid angle subtended by the calculation domain
from the center $O$ is given by
\begin{equation}
        \Omega_\mathrm{b} \simeq 4 \theta_\mathrm{b}^2\;\;\;\mathrm{for}\;\;\theta_\mathrm{b}\ll1,
\end{equation}
the total number of SPH particles can be reduced 
by a factor of $\Omega_\mathrm{b}/4\pi\sim\theta_\mathrm{b}^{2}/\pi$ 
compared with the calculation of the whole shell.

\subsection{Simulation Units}
In this paper, for convenience, the units of the time, length, 
and mass scales are taken to be
\begin{equation}
  t_0 = \sqrt{\frac{3\pi}{32G\rho_\mathrm{E}}} = 
  1.6\;n_\mathrm{E,3}^{-1/2}\;\mathrm{Myr},
  \label{t scale}
\end{equation}
\begin{equation}
  R_0 = \left( \frac{7 c_\mathrm{II} t_0}{\sqrt{12}} \right)^{4/7} R_\mathrm{ST}^{3/7}
  = 5.9\;
  Q_\mathrm{UV,49}^{1/7}\:
  n_\mathrm{E,3}^{-4/7}\;\mathrm{pc},
  \label{r scale}
\end{equation}
and
\begin{equation}
  M_0 = \rho_\mathrm{E} R_0^3
  = 5.0\times10^3\;
  Q_\mathrm{UV,49}^{3/7}\:
  n_\mathrm{E,3}^{-5/7}\;M_\odot,
  \label{m scale}
\end{equation}
respectively, where $Q_\mathrm{UV,49}=Q_\mathrm{UV}/10^{49}\:\mathrm{s}^{-1}$,
and $n_\mathrm{E,3}=n_\mathrm{E}/10^3\:\mathrm{cm^{-3}}$.
The dependence of the expansion law on the parameters $(Q_\mathrm{UV},\:n_\mathrm{E})$ are 
approximately eliminated by using the non-dimensional quantities normalized by 
$t_0$, $R_0$, and $M_0$ (see Figure 1 in Paper I).

\subsection{Initial and Boundary Conditions}\label{sec:ini bou}
It is difficult to resolve 
very thin shell in early evolutionary phase even if 
we calculate the part of the shell (see Figure \ref{fig:region}).
Thus, we use a grid-based 1D hydrodynamical 
code to describe the early evolutionary phase of the HII region
expansion, and we switch to 3D SPH at $t=0.4t_0$.
As a grid-based numerical method, we use 
the 1D spherically symmetric Lagrangian 
Godunov method \citep{vL97}.
The innermost mesh at $R_\mathrm{b}$ is pushed by the interior
pressure given by Equation (\ref{Pext}).
When the 1D simulation reaches $t/t_0=0.4$, 
the radial profiles of physical quantities
are used as the initial condition of the 3D calculations.
We smooth the distribution of the sound speed  across the CD 
in the scale of the smoothing length.
It is assumed that 
the temperature of the hot gas is 64 times as high as 
that of the cold gas.
The specific value of the 
hot gas temperature does not change our conclusion as long as 
it is much larger than the temperature of the cold gas.

We calculate the expansions of HII regions around the 
41$M_{\odot}$ (the high-mass (HM) model) and 19$M_\odot$ (the low-mass (LM) model) stars 
that are embedded by the uniform ambient gas of $n_\mathrm{E}=10^3$ cm$^{-3}$.
Simulation parameters are listed in Table \ref{table:model para}.
The temperature of the cold gas $T_\mathrm{c}$ is assumed to be $10$ K.
The opening angles of simulation domains are set to 
$\theta_\mathrm{b}=2\pi/26$ and $2\pi/14$ for 
the HM and the LM models, respectively, so that 
the calculation domains contain a single wavelength of the most 
unstable mode predicted from the thin-shell linear analysis by \citet{E94}.
The mass of SPH particles are set 
so that the initial thickness is resolved by five SPH particles.
Since the thickness increases with time, 
the relative resolution becomes better as the shell expands.
In the later phase, the thickness is resolved by 
$\sim15$ particles.
The total number of SPH particles for the HM and the LM models are 
$4.00\times10^6$ and $2.26\times10^6$, respectively.

Corrugation-type perturbations are put into the shell at 
the initial state.
We displace the SPH particles so that their densities do not change,  
or ${\bf \nabla}\cdot \Delta {\bf x}_i=0$, 
where $\Delta{\bf x}_i$ is the displacement of the $i$-th particle.
The functional form of the 
displacement of the shell is assumed to be $\propto -\cos (l\phi)$, 
where $\phi = \tan^{-1}(y/x)$. 
Therefore, the shell has the negative displacement at $\phi=0$.
The SPH particles are displaced keeping their velocity, 
the sound speed, and the mass fixed.
We concentrate on the evolution of a single mode in each calculation.
Dependence on $l$ is investigated by many runs.

The moving boundary condition at $r=R_\mathrm{b}(t)$ 
is realized by using ``ghost particles'' \citep{TMS94}.
The time evolution of $R_\mathrm{b}(t)$ is obtained by the results of 
the 1D simulations.
At the four surface areas ($y=\pm x\tan\theta_\mathrm{b}$, $z
=\pm x\tan\theta_\mathrm{b}$)
in the pyramid-shaped calculation domain (see Figure \ref{fig:region}),
the rotational periodic boundary condition is imposed.

\begin{figure}[htpb]
        \begin{center}
            \includegraphics[width=8cm]{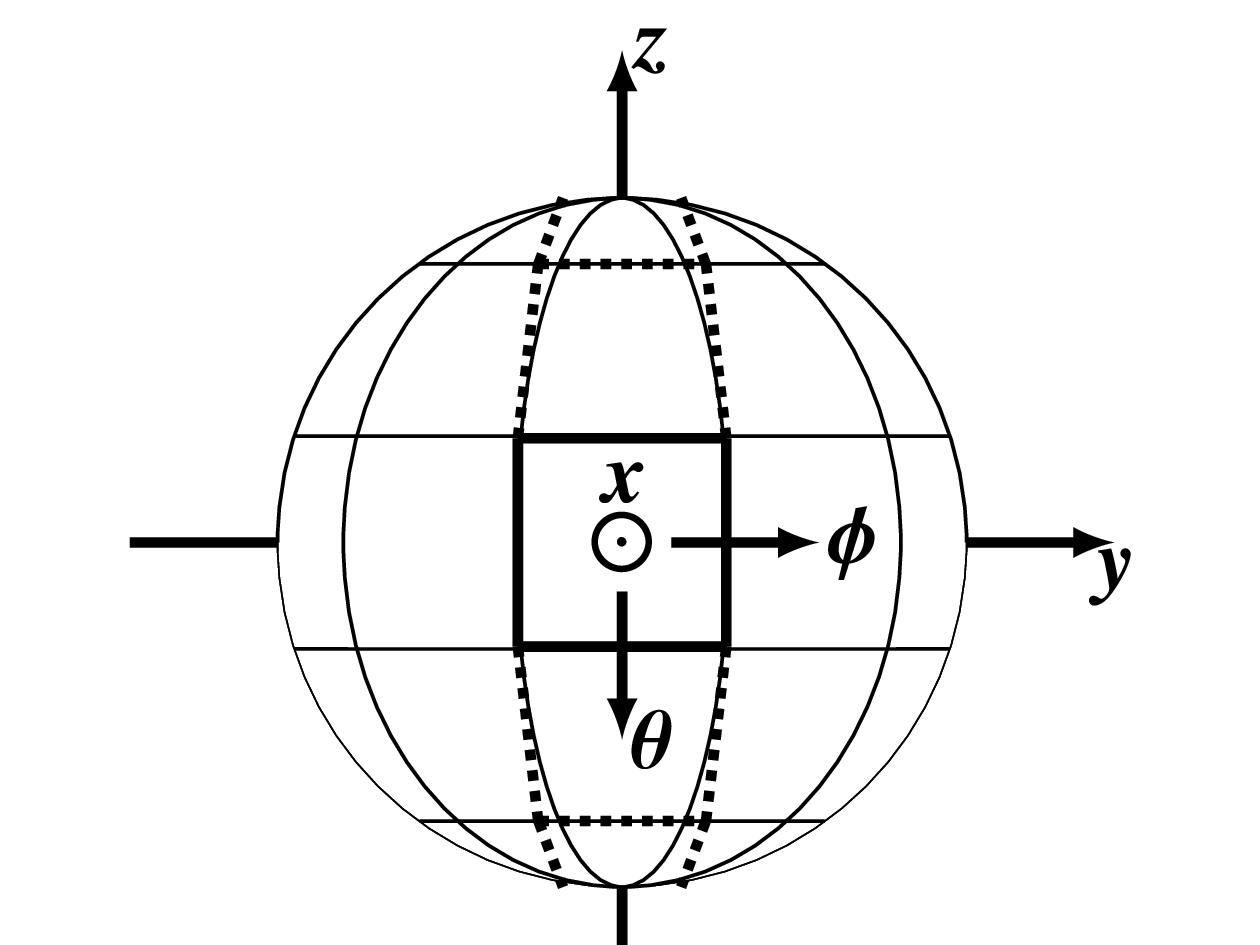}
        \end{center}
\caption{
Schematic picture viewed from the $x$-direction for the case with $\theta_\mathrm{b}=2\pi/10$. 
The latitude and longitude lines are plotted at interval of $\theta_\mathrm{b}$.
The region is surrounded by the thick solid lines represents the calculation domain 
(see Figure \ref{fig:region}). 
The dashed regions correspond to rotated domains in the $\theta$-direction.
}
\label{fig:grav region}
\end{figure}

Since the gravitational force is a long-range force, 
we have to take into account the particles in the whole solid angle. 
However, we only have information in the part of the solid angle 
as shown in Section \ref{sec:domain}.
In this paper, the gravitational force is evaluated in the following method.
Figure \ref{fig:grav region} shows the schematic picture of the gas sphere viewed from the $x$-direction for the 
case with $\theta_\mathrm{b}=2\pi/10$.
The latitude and longitude lines are plotted at interval of $\theta_\mathrm{b}$.
The region is surrounded by the thick solid lines represents the calculation domain (see Figure \ref{fig:region}). 
We rotate the calculation domain $i\theta_\mathrm{b}$ degrees in the $\theta$-direction, where 
$i=-\pi/(2\theta_\mathrm{b}), \ldots, \pi/(2\theta_\mathrm{b})$.
At each rotation step in the $\theta$-direction, we rotate the domain $k\theta_\mathrm{b}$ degrees
in the $\phi$-direction, where $k=1,\ldots,2\pi/\theta_\mathrm{b}$.
After that, the information in the whole solid angle can be obtained.
However, for $i\ne0$, the rotated domains in the $\phi$-direction have overlaps.
This comes from the fact that the pyramid cannot fill 3D space.
For example, the dashed regions in Figure \ref{fig:grav region} correspond to the rotated domains in 
the $\theta$-direction.
We remove the overlapped SPH particles with $|\tan^{-1}(y/x)|>\theta_\mathrm{b}$ from 
the rotated domains. In the simulation, we rotate not SPH particles but the tree structure, and 
calculate the gravitational force from the rotated tree structures at each rotation step.

By this method, supposed particle distribution in the whole solid angular $4\pi$ is 
not cyclic completely. However, the effect of the approximation is negligible since 
$\theta_\mathrm{b}$ is very small in this paper.
For $i=\pm1$ (near the equatorial plane), the number of removed SPH particle is 
negligible compared with the total number.
On the other hand, for large $|i|$ (near the poles), the fraction of removed SPH particles becomes large. 
However, since the rotated domain is very far from the calculation domain, 
the detail particle distribution is not important.

In this paper, we neglect the gravitational force outside the SF
to prevent the ambient gas from collapsing toward the center.
We find the radius $R_\mathrm{grav}$ of the most distant particle from the center that 
has density $\rho>\rho_\mathrm{th}$, where $\rho_\mathrm{th}$ is a threshold density \citep{Betal09}.
In our simulations, we adopt $\rho_\mathrm{th}=1.1\rho_\mathrm{E}$.
Only SPH particles within $R_\mathrm{grav}$ are assumed to feel the gravitational force that is calculated
above method. This means that there is the discontinuity in the gravitational force at 
the $R_\mathrm{grav}$. In the Appendix, we investigate the effect of the discontinuity, and 
confirm that it does not influence our results as long as $R_\mathrm{grav}$ is inside the shock transition layer. 
The main reason is that the pressure suddenly changes in the shock transition layer, suggesting that
the pressure gradient is much larger than the jump in the gravitational force.

\begin{table*}
\centering
 \begin{tabular}{cccccc}
        \hline
   Model & $M_*(M_\odot)^a$ & $R_\mathrm{ST}(\mathrm{pc})$ & $R_0(\mathrm{pc})$ & $T_0$(Myr) & $M_0(M_\odot)$  \\
        \hline
        \hline
     High mass (HM)& 41 & 0.56 & 5.5 & 1.6 & $4.1\times10^3$ \\
        \hline
     Low mass (LM)& 19  & 0.25 & 3.9 & 1.6 & $1.46\times10^3$ \\
        \hline
   \end{tabular}
     {\footnotesize
     \begin{flushleft}
     $^a$ Mass of the star \\
     \end{flushleft}
     }
   \caption{Parameters of Our Models}
\label{table:model para}
\end{table*}

\subsection{Measures of Fluctuations}\label{sec:focus}
In order to evaluate growth of perturbations in the SPH calculations, 
we introduce indicators for fluctuations of density and the position of the CD.
The solid angle $\Omega_\mathrm{b}$ is divided into $N_\Omega$ cells, and 
the direction of the $i$-th cell is defined by the 
unit vector ${\bf n}_i$ $(1\le i\le N_\Omega)$.
The radial profiles of the physical quantities along ${\bf n}_i$ 
are obtained from the SPH calculations.
Here, we define the angle-dependent 
maximum density $\rho_\mathrm{max}({\bf n}_i)$ along ${\bf n}_i$.
The position of the CD along ${\bf n}_i$ is determined as the 
position where the temperature coincides with a critical value
that is set to $\sqrt{2}T_\mathrm{c}$.
The specific choice of the critical 
value is not important
in our results.
We average $\rho_\mathrm{max}({\bf n}_i)$ and $R_\mathrm{CD}({\bf n}_i)$
over all directions. 
The mean value of $Q=(\rho_\mathrm{max}$, $R_\mathrm{CD})$  
is defined by
\begin{equation}
	\langle Q \rangle = \frac{1}{N_\Omega}\sum_{i} Q({\bf n}_i).
      \label{Qave}
\end{equation}
As the indicators of the fluctuations, 
we evaluate the dispersions of these quantities 
\begin{equation}
  \Delta Q \equiv \sqrt{\frac{1}{N_\Omega}\sum_{i} 
	\left\{ Q({\bf n}_i) - \langle Q\rangle\right\}^2}.
      \label{delta Q}
\end{equation}

\subsection{Unperturbed Evolution}\label{sec:without per}
\begin{figure}[htpb]
        \begin{center}
            \includegraphics[width=8cm]{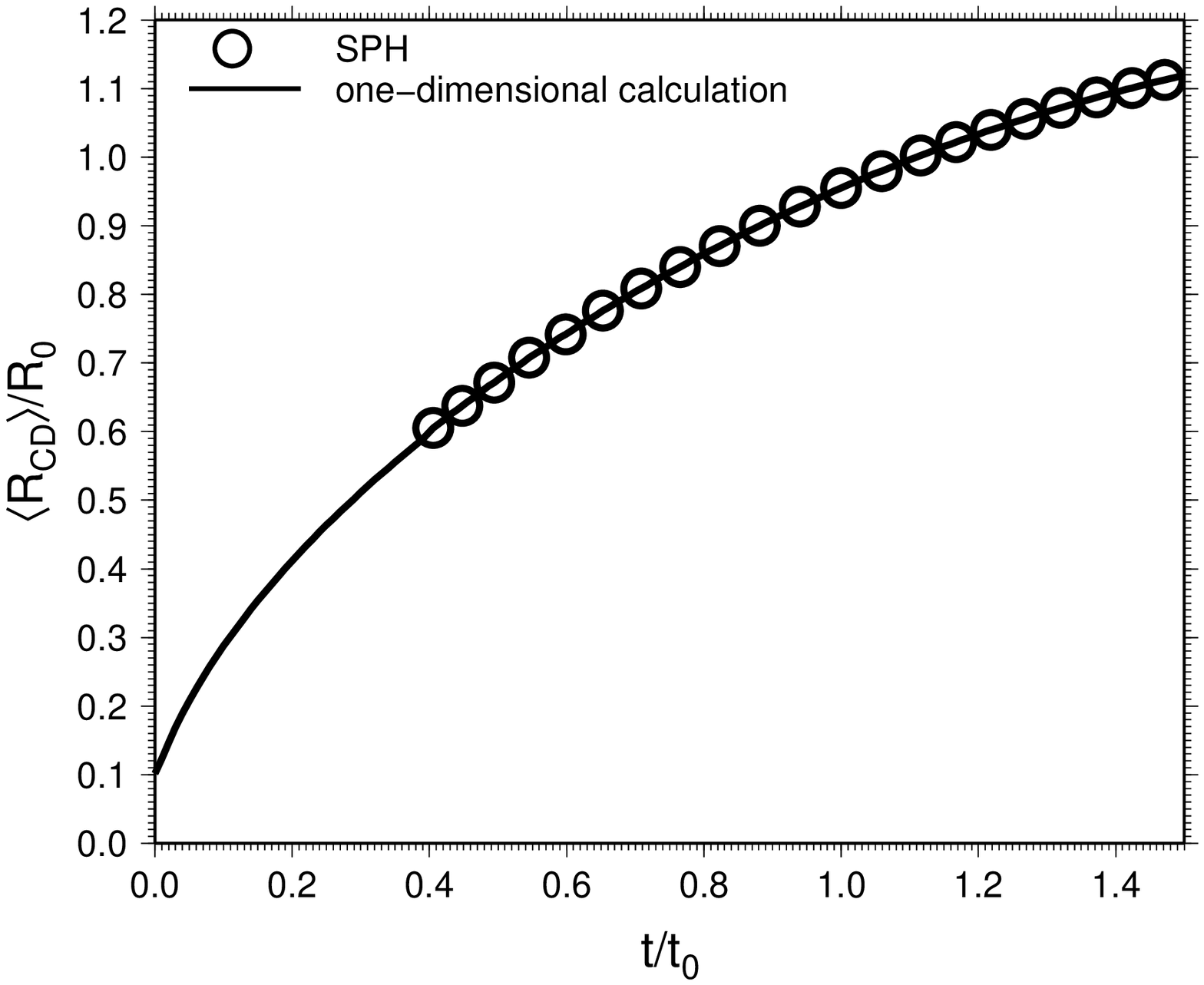}
        \end{center}
\caption{
Time evolution of $\langle R_\mathrm{CD}\rangle$. The open circles indicate
the result of the SPH calculation. The solid line
indicates the result of the 1D calculation.
}
\label{fig:surpos}
\end{figure}

As a test calculation, we present the evolution of the shell 
without initial perturbations in the HM model.
Figure \ref{fig:surpos} shows the time evolution 
of $\langle R_\mathrm{CD}\rangle$ (Equation (\ref{Qave}))
evaluated from the SPH simulation (the open circles). 
For comparison, we plot the result of the 1D simulation by the solid line.
One can see that the SPH simulation can reproduce the results 
of the 1D calculation very well.

\begin{figure}[htpb]
        \begin{center}
            \includegraphics[width=8cm]{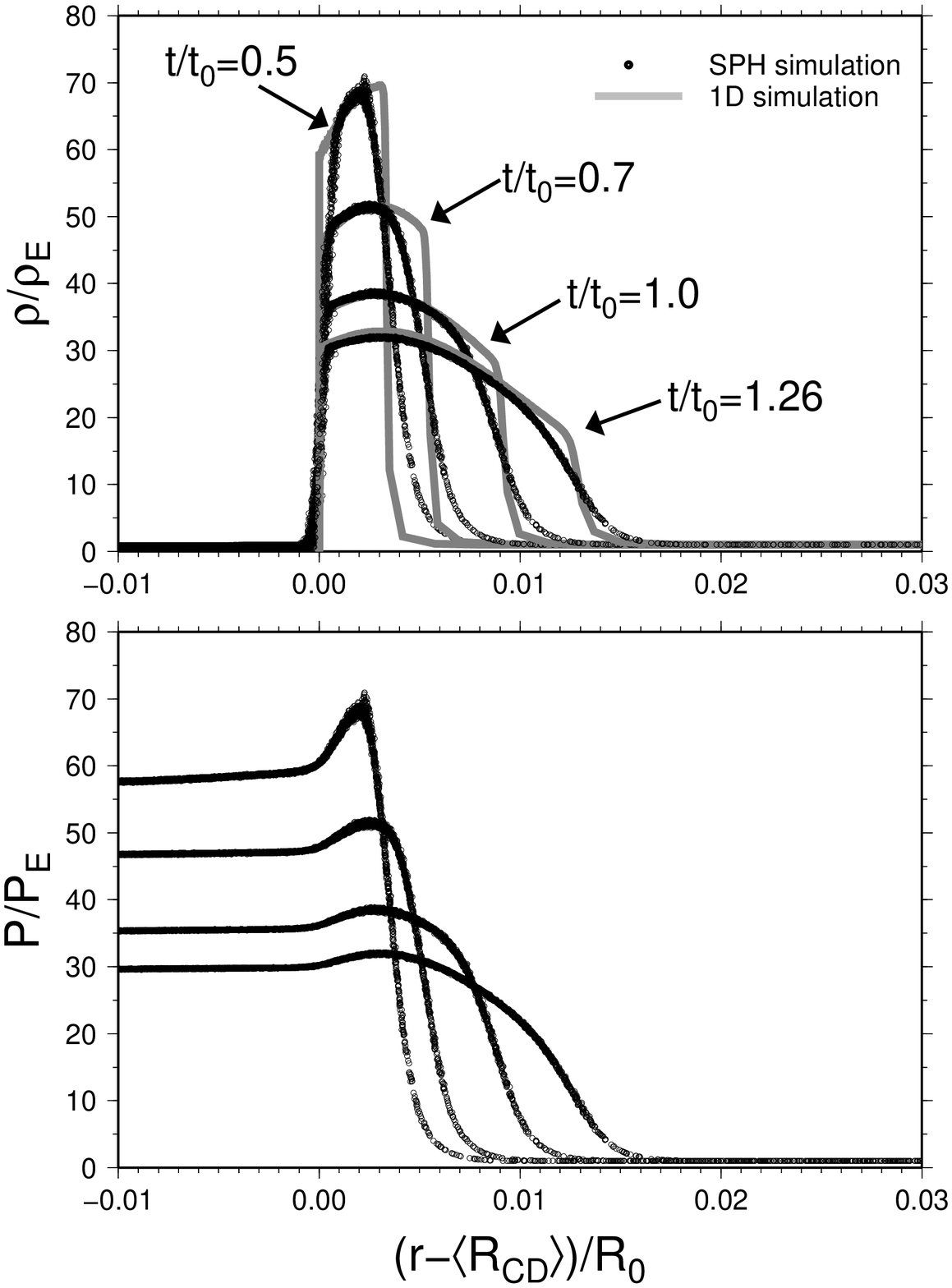}
        \end{center}
\caption{
Snapshots of density (the upper panel) and pressure 
(the lower panel) profiles for
$t/t_0=0.5$, 0.7, 1.0, and 1.26.
Only SPH particles in $|y|/R_0<0.05$ and $|z|/R_0<0.05$ are plotted by 
the dots.
The abscissae are the distance from the CD.
The thick gray lines in the upper panel indicate the results of the 
1D simulation.
}
\label{fig:1dim}
\end{figure}

Figures \ref{fig:1dim} show snapshots of density (the upper panel) 
and pressure (the lower panel) profiles
for $t/t_0=0.5$, 0.7, 1.0, and 1.26.
The abscissae indicate the distance from the CD at each time.
Only SPH particles in $|y|/R_0<0.05$ and $|z|/R_0<0.05$ are plotted 
by the dots.
For comparison, the density profiles in the 1D simulation are
superimposed by the thick gray lines in the upper panel.
It is seen that the result of the SPH calculation agrees with that of 
the 1D simulation very well.
Our 3D simulation clearly represents the structure and evolution of 
the shell although it is very thin. 

\begin{figure}[htpb]
        \begin{center}
            \includegraphics[width=8cm]{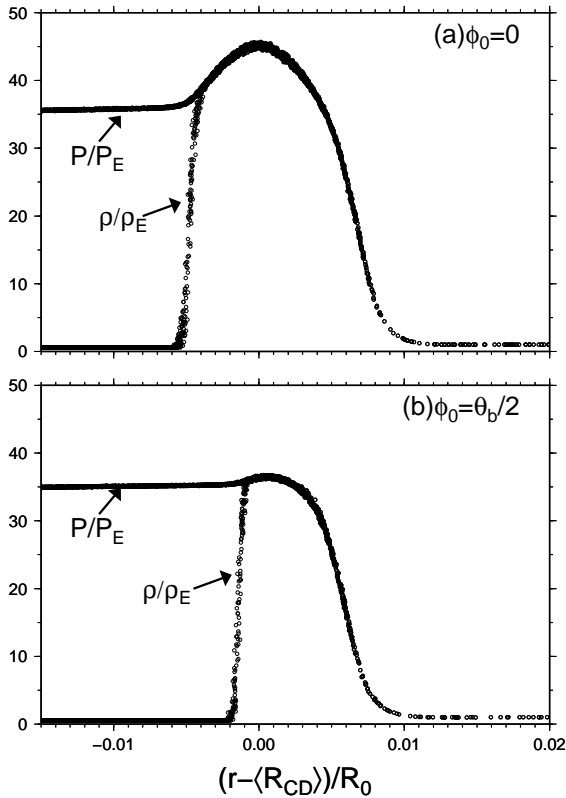}
        \end{center}
\caption{
Snapshots of the density and pressure profiles for $t/t_0=1.0$ and $l=52$.
The abscissae are the distance from the average position of the CD.
SPH particles in $\phi_0-\Delta \phi<\phi<\phi_0+\Delta \phi$ and $|z|/R_0<0.05$ are plotted, 
where $\phi_0=$(a)0, (b)$\theta_\mathrm{b}/2$, and $\Delta \phi=0.001\theta_\mathrm{b}$.
}
\label{fig:1dim per}
\end{figure}

Note that the pressure profiles in the lower panel of Figure \ref{fig:1dim} 
are smooth at the CD $(r-\langle R_\mathrm{CD}\rangle=0)$ thanks to 
the modified Godunov SPH shown in Section \ref{sec:sph}. 
In the standard SPH, a wiggle in pressure profile appears at the CD.
In order to see whether the pressure profile is smooth even when perturbation is added, 
the density and pressure profiles for $t=1.0$ and $l=52$ are plotted in 
Figure \ref{fig:1dim per}.
The detailed investigation of the results with perturbation is shown in Section \ref{sec:sph result}.
The abscissa is the distance from $\langle R_\mathrm{CD}\rangle$.
We plot SPH particles in $\phi_0-\Delta \phi<\phi<\phi_0+\Delta \phi$ and $|z|/R_0<0.05$, where 
$\Delta \phi=0.001\theta_\mathrm{b}$.
Figures \ref{fig:1dim per}(a) and (b) correspond to $\phi=0$ where $\delta \rho$ 
has the maximum value and $\phi_0=\theta_\mathrm{b}/2$ where $\delta \rho$ has the minimum value, 
respectively. One can clearly see that the pressure profile is smooth at the CD even with perturbation.
\section{Prediction by Linear Analysis}\label{sec:review}
\citet{IIT11} performed a linear analysis taking into account the thickness 
of the shell as in Figure \ref{fig:1dim} 
and imposing the approximate SF and the CD boundary conditions.
They neglected evolutionary effect, i.e., expansion and mass accretion through
the SF. At each instant of time, they solved eigen-value problem
and obtained the growth rate $\Gamma(l,t)$ as 
a function of the time and the angular wavenumber, where 
$\Gamma$ is the imaginary part of $\omega(k,t)$ in Paper I.
The angular wavenumber $l$ can be expressed as $2\pi R_\mathrm{s}/k$ by using 
the wavenumber $k$ in Paper I, where $R_\mathrm{s}$ is the shell radius.

\subsection{Time Evolution and Scaling Law of Density Profile}\label{sec:den pro}
The density profile of the decelerating shell is asymmetric with respect to the density peak.
\citet{IIT11} developed a semi-analytic method for deriving the time evolution of the 
density profile by assuming that the shell reaches the hydrostatic equilibrium 
among the pressure gradient, the inertia force owing to the deceleration, and 
the radial gravitational force toward the center \citep{WF02}.
\citet{IIT11} found that the time evolution of the density profile 
can be divided into three evolutionary phases, deceleration-dominated, intermediate, 
and self-gravity-dominated phases (see Figure 2 in Paper I).
In the early phase (deceleration-dominated phase), 
the density peak is at the SF because the inertia force is larger than the gravitational force. 
As the shell expands, the inertia force decreases while the gravitational force increases.
Thus, the density peak moves from the SF toward the CD as shown in Figure \ref{fig:1dim}.
In the intermediate phase, the density peak is closer to the SF than the CD.
In the later phase (self-gravity-dominated phase), the density peak is closer to the CD than
the SF.

It is also found that the density profiles
for various sets of $(n_\mathrm{E},\;Q_\mathrm{UV})$ 
are characterized by a single parameter, that is the 
typical Mach number,
\begin{equation}
        {\cal M}_0
        = \frac{4}{7}\frac{R_0}{c_\mathrm{s} t_0}= 7\;
      Q_\mathrm{UV,49}^{1/7}\:
      T_\mathrm{c,10}^{-1/2}\:
      n_\mathrm{E,3}^{-1/14},
  \label{mach}
\end{equation}
where $T_\mathrm{c,10}=T_\mathrm{c}/10$ K.

They found the development of the GI is strongly influenced by 
two effects, asymmetry of the density profile and boundary conditions.
\subsection{Asymmetry of the Density Profile}\label{sec:asym}
The asymmetry of the density profile strongly influences the GI, especially, 
in the self-gravity-dominated phase (for example, the shell at $t/t_0=1.26$ in Figure \ref{fig:1dim}). 
In this later phase, the distance from the density peak to the SF is larger than $H_0$, where
$H_0=c_\mathrm{s}/\sqrt{2\pi G\rho_{00}}$ 
is the scale height of the shell and $\rho_{00}$ is the peak density.
On the other hand, the distance from the density peak to the CD is smaller than $H_0$.
The gas tends to collect toward the density peak because the unperturbed gravitational 
potential has the minimum value there.
The gas near the SF can collapse toward the density peak because the sound wave cannot 
travel from the density peak to the SF. This mode is ``compressible mode''.
On the other hand, the gas near the CD cannot collapse to 
the peak. However, the GI can proceed even there through the deformation of the CD 
that makes the gravitational potential deeper.
This mode is ``incompressible mode'' \citep{EE78, LP93}.
Therefore, the GI of the shell in the later phase has properties of 
compressible and incompressible modes.

Moreover, \citet{IIT11} found that the growth rate is enhanced compared with symmetric case through 
the combination of the GI and the Rayleigh-Taylor instability.
\subsection{Boundary Conditions}\label{sec:boundary}
Expanding shell is confined by the SF on the leading surface and 
by the CD on the trailing surface.
Dispersion relations of \citet[][E94]{E94} and \citet{Wetal94} are essentially 
based on the dispersion relation of the layer confined by the SF on both surfaces.
The shock-confined layer is stabilized 
by the tangential flow generated by obliqueness of the SF compared with
the layer confined by thermal pressure of hot gas. 
Therefore, the linear analysis in Paper I taking into account SF+CD boundary conditions 
gives larger growth rate compared with E94.

In summary, as described in Sections \ref{sec:asym} and \ref{sec:boundary}, 
Paper I predicts that 
the GI begins to grow earlier than the prediction from E94.
The development of the GI in the later phase is accompanied by 
the significant deformation of the CD.

\section{Results}\label{sec:sph result}

\subsection{The Case with High-mass Central Star}
\begin{figure}[htpb]
        \begin{center}
            \includegraphics[width=8cm]{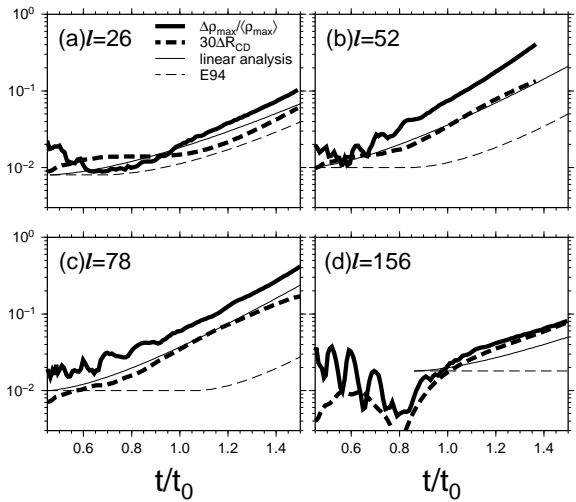}
        \end{center}
\caption{
Time evolution of perturbations for the HM model. 
Each panel corresponds to $l=$ (a) 26, (b) 52, (c) 78, and (d) 156.
The thick solid and the thick dashed lines represent the density perturbation
$\Delta \rho_\mathrm{max}/\langle\rho_\mathrm{max}\rangle$ and the deformation of the CD
$30\times\Delta R_\mathrm{CD}$, respectively.
The thin solid lines correspond to the prediction from Paper I.
The thin dashed lines correspond to the prediction based on \citet{E94}.
}
\label{fig:model1_growth}
\end{figure}

First, we present the results of the HM model with perturbations.
We calculate four runs for the case with $l=26$, 52, 78, and 156.
Figures \ref{fig:model1_growth}(a), (b), (c), and (d)
represent the time evolution of the perturbation amplitude
for the wavenumber $l=26$, 52, 78, and 156, respectively.
The thick solid and the thick dashed lines correspond to  
the density perturbation $\Delta \rho_\mathrm{max}/\langle\rho_\mathrm{max}\rangle$ and 
the deformation of the CD
$30\times\Delta R_\mathrm{CD}$, respectively.
In each panel, $\Delta \rho_\mathrm{max}/\langle\rho_\mathrm{max}\rangle$ and 
$\Delta R_\mathrm{CD}$ grow in the similar way.
Perturbations with $l=52-78$ grow with the largest growth rate.
For larger angular wavenumber $l=156$ and smaller 
angular wavenumber $l=26$, they grow more slowly.
The prediction from E94 is also superimposed by the thin dashed line in each panel.
It is found that in our 3D results, perturbations begin to grow earlier than the 
prediction of E94.  

Let us compare the results of SPH simulations with the linear analysis in Paper I.
However, Paper I neglected the evolutionary effects, i.e., expansion of the shell 
and the mass accretion through the SF.
To include these effects approximately, we modify the instantaneous growth rate from 
$\Gamma$ to $\Gamma_\mathrm{evo}$ as follows:
\begin{equation}
	\Gamma_\mathrm{evo}(l,t) \equiv -3\frac{V_\mathrm{s}(t)}{R_\mathrm{s}(t)} 
      + \sqrt{\left( \frac{V_\mathrm{s}(t)}{R_\mathrm{s}(t)} \right)^2 + \Gamma(l,t)^2},
      \label{mod}
\end{equation}
where $V_\mathrm{s}$ is the velocity of the shell.
This modification is inspired by E94. 
The time evolution of $R_\mathrm{s}(t)$ and $V_\mathrm{s}(t)$ is determined by the thin-shell model 
shown in Section 2 in Paper I.
One can see that the terms of $V_\mathrm{s}/R_\mathrm{s}$ 
correspond to evolutionary effect of the shell that stabilizes the GI.
The prediction based on 
Paper I is superimposed by the thin solid line in each panel of
Figure \ref{fig:model1_growth}.  It is evaluated by 
$\exp\left[\int_t \Gamma_\mathrm{evo}(l,t') \mathrm{d} t'\right]$.
From Figure \ref{fig:model1_growth}, 
one can see that the linear analysis of Paper I describes the growth of perturbations very well.

\begin{figure}[htpb]
        \begin{center}
         \includegraphics[width=8cm]{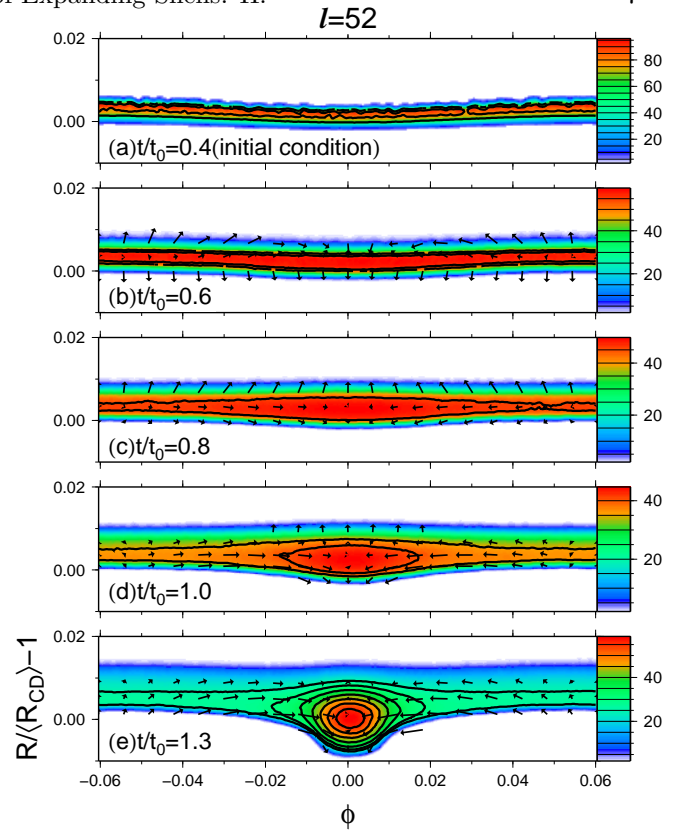} 
        \end{center}
      \caption{
      Cross sections through the $z=0$ plane for $l=52$.
      Each panel shows the shell at 
      $t/t_0=(a)0.4$, (b)0.6, (c)0.8, (d)1.0, and (d)1.3, respectively.
      The color scale represents the density, and arrows are velocities relative to 
      the fluid at the density peak.
      The abscissa and the ordinate axes correspond to the azimuthal angle $\phi=\tan^{-1}(y/x)$,
      and $R/\langle R_\mathrm{CD}\rangle-1$, respectively. 
     }
 \label{fig:den_l52}
\end{figure}

Figure \ref{fig:den_l52} shows the time sequence of 
cross sections through the $z=0$ plane for $l=52$.  
The color scale indicates density, and 
the arrows represent relative velocity to 
the fluid with the maximum density.  
The abscissa and the ordinate axes correspond to the 
azimuthal angle $\phi$,
and the distance from the CD divided by $\langle R_\mathrm{CD}\rangle$, 
respectively.
Figure \ref{fig:den_l52}(a) represents the initial condition
where the shell around $\phi=0$ is displaced to the negative radial direction.
Figure \ref{fig:den_l52}(b) shows that the tangential flow 
is generated by the obliqueness of the SF, and 
collects the gas around $\phi=0$.
As a result, enhanced pressure pushes the SF outward in the 
radial direction (see Figure \ref{fig:den_l52}(c)).
The gravitational potential around $\phi=0$ becomes deep, leading to 
gas accumulation (see Figure \ref{fig:den_l52}(d)).
This mode of the GI is similar to the incompressible mode in the sense that
the deformation develops at the boundary \citep{EE78,LP93}.
In the later phase (Figure \ref{fig:den_l52}(e)), 
the CD highly deforms while the SF is almost flat. 
This is consistent with prediction of Paper I (see Section \ref{sec:review}).
Figure \ref{fig:den_l52}(e) is very similar to Figure 11 in Paper I that shows the 
eigenfunction in parameters of the HM model at $t/t_0=1.3$ for $l=52$.

\begin{figure}[htpb]
       \begin{center}
         \includegraphics[width=8cm]{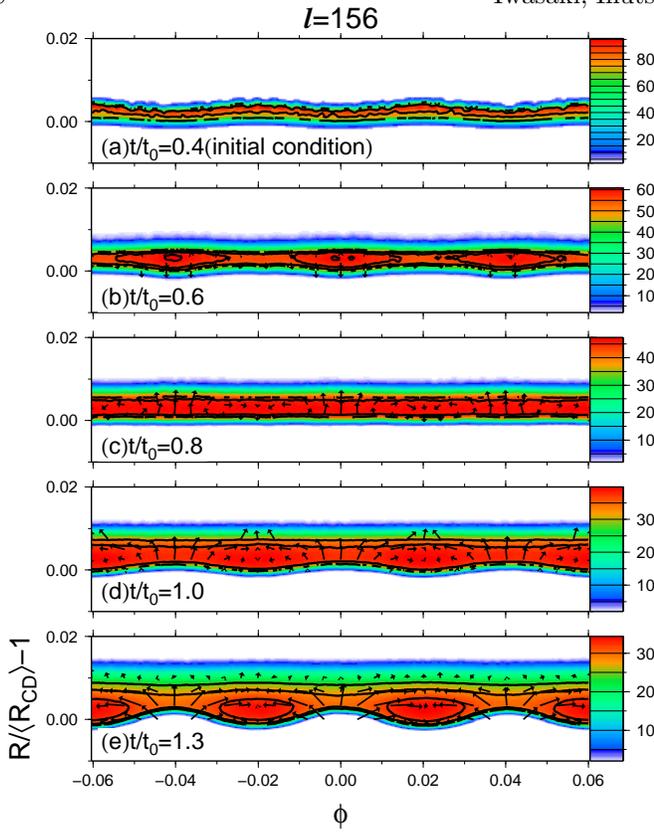} 
       \end{center}
      \caption{
      Same as Figure \ref{fig:den_l52} but for $l=156$.
     }
 \label{fig:den_l156}
\end{figure}

Figure \ref{fig:den_l156} is the same as Figure \ref{fig:den_l52} but 
for larger wavenumber case, $l=156$.
The initial state is shown in Figure \ref{fig:den_l156}(a).
Because of large wavenumber, the tangential flow collects the gas 
more quickly in Figure \ref{fig:den_l156}(b) that is similar to Figure 
\ref{fig:den_l52}(c).
In this moment, since the SF at $\phi=0$ deforms to the positive radial direction, 
the gas moves away from $\phi=0$ 
by the tangential flow. As a result, at $t/t_0=0.8$ (Figure \ref{fig:den_l156}(c)), 
the SF and the CD are almost flat while the velocity field exists inside the shell.
The density distribution at $t/t_0=1.0$ 
(Figure \ref{fig:den_l156}(d)) 
has the opposite phase compared to the initial state.
In this moment, the perturbation begins to grow but the growth rate is smaller than that 
for $l=52$.
The CD deforms largely at the final state in both of 
Figures \ref{fig:den_l52} and \ref{fig:den_l156}.
In this sense, the behavior of the GI is similar to that for $l=52$. 

\subsection{The Case with Low Mass Central Star}
\begin{figure}[htpb]
        \begin{center}
            \includegraphics[width=8cm]{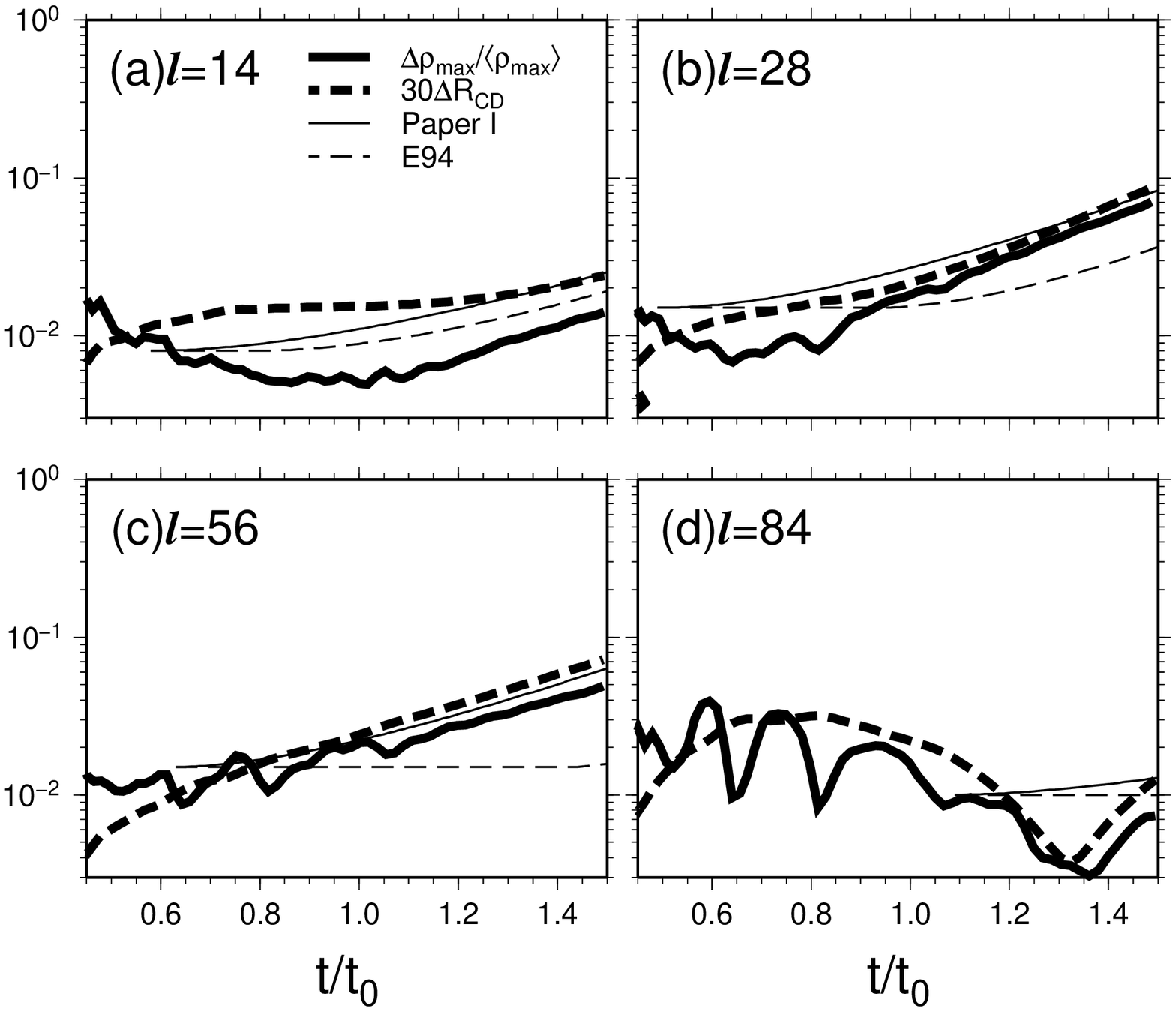}
        \end{center}
\caption{
The same as Figure \ref{fig:model1_growth} but for the LM model.
Each panel corresponds to $l=$(a)14, (b)28, (c)56, and (d)84. 
}
\label{fig:model2_growth}
\end{figure}

\begin{figure}[htpb]
        \begin{center}
            \includegraphics[width=8cm]{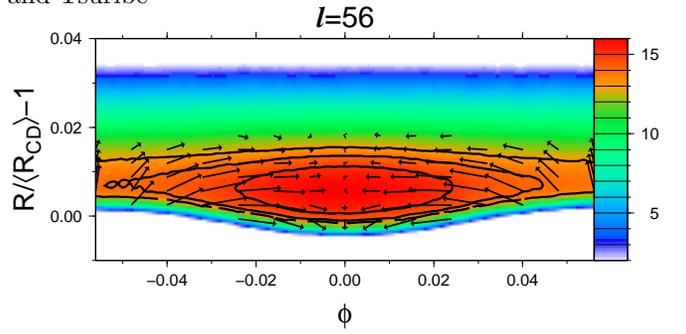}
        \end{center}
\caption{
Cross section through $z=0$ plane for $l=56$ at $t/t_0=1.5$ for the 
LM model.
The color scale represents the density, and the arrows 
show velocities relative to 
the fluid at the density peak.
The abscissa and ordinate axes are the same as Figure \ref{fig:den_l52}.
}
\label{fig:den_l56}
\end{figure}

Next, we present results of the LM model.
Because the central star is less energetic, 
the Mach number of the expanding shell is smaller than that in the 
HM model,
suggesting that the shell is geometrically thicker.
Therefore, perturbations are expected to grow more slowly than the HM model.
This feature is confirmed in Figure 
\ref{fig:model2_growth} that is similar to
Figure \ref{fig:model1_growth}.
The most unstable mode is found in smaller wavenumber $l\sim28$ and 
the growth rate is smaller than the HM model.
We found that perturbations begin to grow earlier than the 
prediction of E94 also in this case.
The linear analysis in Paper I describes the GI very well.
Figure \ref{fig:den_l56} 
shows the cross section through $z=0$ for $l=56$ at $t/t_0=1.5$.
The significant deformation of the CD is seen as in the HM model.

\section{Discussion}\label{sec:discuss}
\subsection{Growth Rate of Gravitational Instability}
\begin{figure*}[htpb]
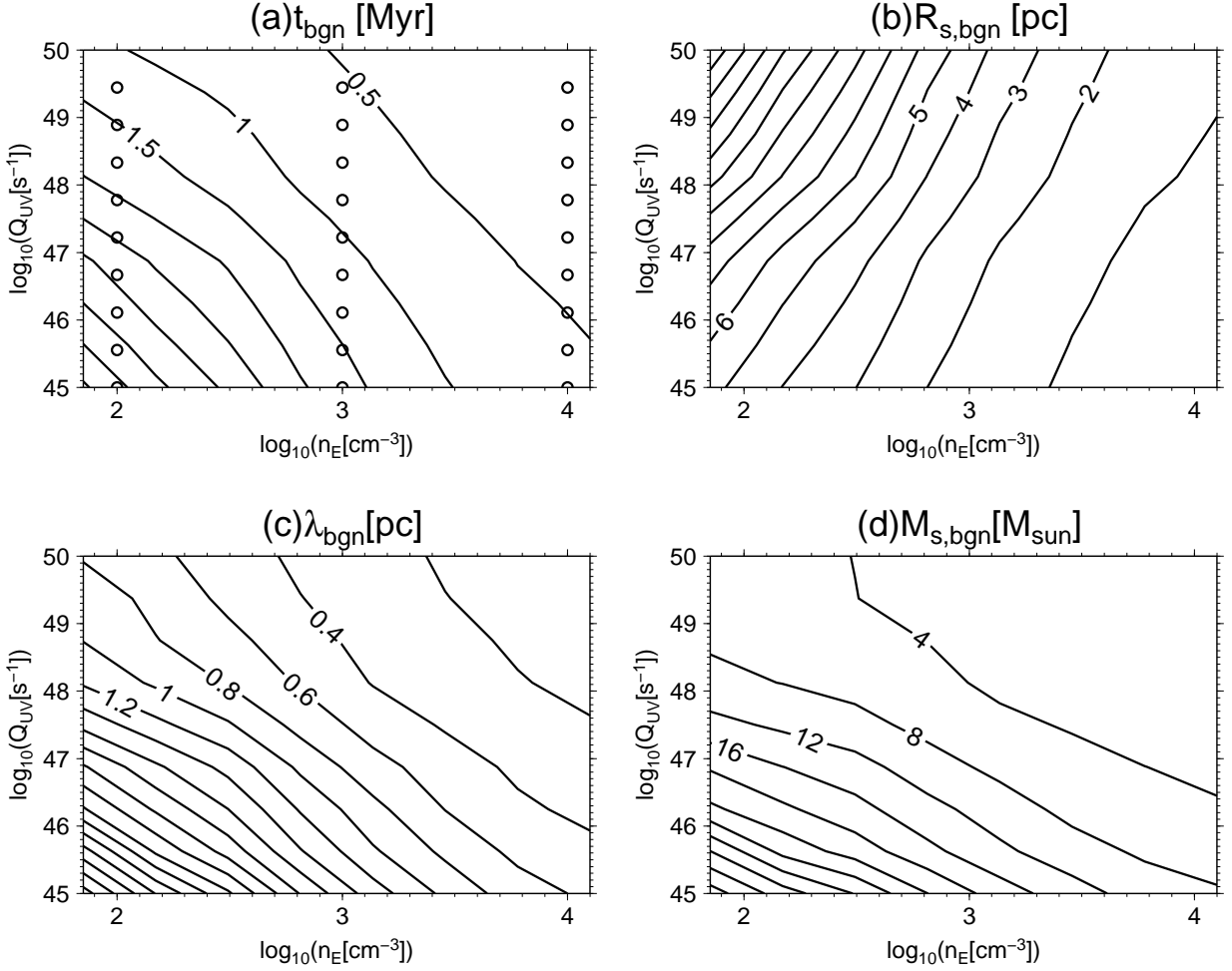

     \begin{center}
        \begin{tabular}{cc}
         \includegraphics[width=8cm]{f12a.eps}
         &
         \includegraphics[width=8cm]{f12b.eps}
        \end{tabular}
     \end{center}
\caption{
Contour maps of (a)$t_\mathrm{bgn}$, (b)$R_\mathrm{s,bgn}$,
(c)$\lambda_\mathrm{bgn}$, and (d)$M_\mathrm{bgn}$
in the $(\log_{10}n_\mathrm{E},\;\log_{10}Q_\mathrm{UV})$ plane.
The open circles correspond to the parameters that are used
in Figure \ref{fig:begin cs}.
}
\label{fig:begin}
\end{figure*}

We presented only two examples of the HM and the LM models 
in Section \ref{sec:sph result}.
In this section, we predict the GI in the shells with various parameters 
$(n_\mathrm{E},\;Q_\mathrm{UV},\;T_\mathrm{c})$ by using the 
results of the linear analysis presented in Paper I.
As shown in Figures \ref{fig:model1_growth} and \ref{fig:model2_growth}, 
it is found that the results of the SPH simulations are well described by the 
growth rate, $\Gamma_\mathrm{evo}$.
Using this growth rate, one can predict the time $t_\mathrm{bgn}$ when 
the GI begins to grow by the condition $\Gamma_\mathrm{evo}(l_\mathrm{bgn},\;
t_\mathrm{bgn})=0$, where $l_\mathrm{bgn}$ is the angular wavenumber of the 
most unstable mode at $t=t_\mathrm{bgn}$.
This condition is rewritten as 
\begin{equation}
        \sqrt{8}\frac{V_\mathrm{s}}{R_\mathrm{s}} = 
        \Gamma(l_\mathrm{bgn},\:t_\mathrm{bgn}),
        \label{frag criterion}
\end{equation}
where the left-hand side corresponds to 
the evolutionary rate owing to expansion and mass accretion,
the right-hand side corresponds to the growth rate of the GI.
At early phase, the evolutionary rate is larger than the growth rate of the GI,
indicating that the shell is stable.
As the shell expands, $V_\mathrm{s}/R_\mathrm{s}$ decreases while $\Gamma$ increases.
Therefore, the GI begins to grow when the growth rate of the GI is larger than
the evolutionary rate.
The radius of the shell at $t=t_\mathrm{bgn}$  is defined by $R_\mathrm{s,bgn}$.

Figure \ref{fig:begin}(a), (b), (c), and (d) show 
contour maps of $t_\mathrm{bgn}$, $R_\mathrm{s,bgn}$, 
$\lambda_\mathrm{max}\equiv2\pi R_\mathrm{s,bgn}/l_\mathrm{bgn}$, and 
$M_\mathrm{bgn}\equiv\Sigma(t=t_\mathrm{bgn}) 
\pi (\lambda_\mathrm{bgn}/2)^2$ in the $(n_\mathrm{E},\;Q_\mathrm{UV})$ plane derived 
by using the linear analysis, respectively.
Here, $T_\mathrm{c}$ is assumed to be 10 K.
From Figures \ref{fig:begin}, one can see that the 
$t_\mathrm{bgn}$, $R_\mathrm{s,bgn}$, $\lambda_\mathrm{bgn}$, 
and $M_\mathrm{bgn}$ decrease with increasing $n_\mathrm{E}$.
This can be understood by scaling of 
the typical number density of the shell, $n_\mathrm{sh}=n_\mathrm{E} {\cal M}_0^2$ that 
corresponds to the shocked gas density.
From Equation (\ref{mach}), we have $n_\mathrm{sh}\propto n_\mathrm{E}^{6/7}$.
Therefore, for larger $n_\mathrm{E}$, the shell becomes denser and unstable earlier 
(Figure \ref{fig:begin}(a)),
and the most unstable scale is smaller (Figures \ref{fig:begin}(c) and (d)).
Next, let us see the dependence of $t_\mathrm{bgn}$, $R_\mathrm{s,bgn}$, $\lambda_\mathrm{bgn}$, 
and $M_\mathrm{bgn}$ on $Q_\mathrm{UV}$.
Since the central star is more energetic for larger $Q_\mathrm{UV}$, 
the Mach number of the expanding shell $\left({\cal M}_0\propto Q_\mathrm{UV}^{1/7}\right)$
is larger, indicating that 
the shell is denser and is more unstable. 
Therefore, $t_\mathrm{bgn}$, $\lambda_\mathrm{bgn}$, and
$M_\mathrm{bgn}$ decreases with increasing $Q_\mathrm{UV}$.
On the other hand, $R_\mathrm{s,bgn}$ increases with $Q_\mathrm{UV}$ because 
the expansion velocity increases with $Q_\mathrm{UV}$.

\begin{figure}[htpb]
      \begin{center}
         \includegraphics[width=8cm]{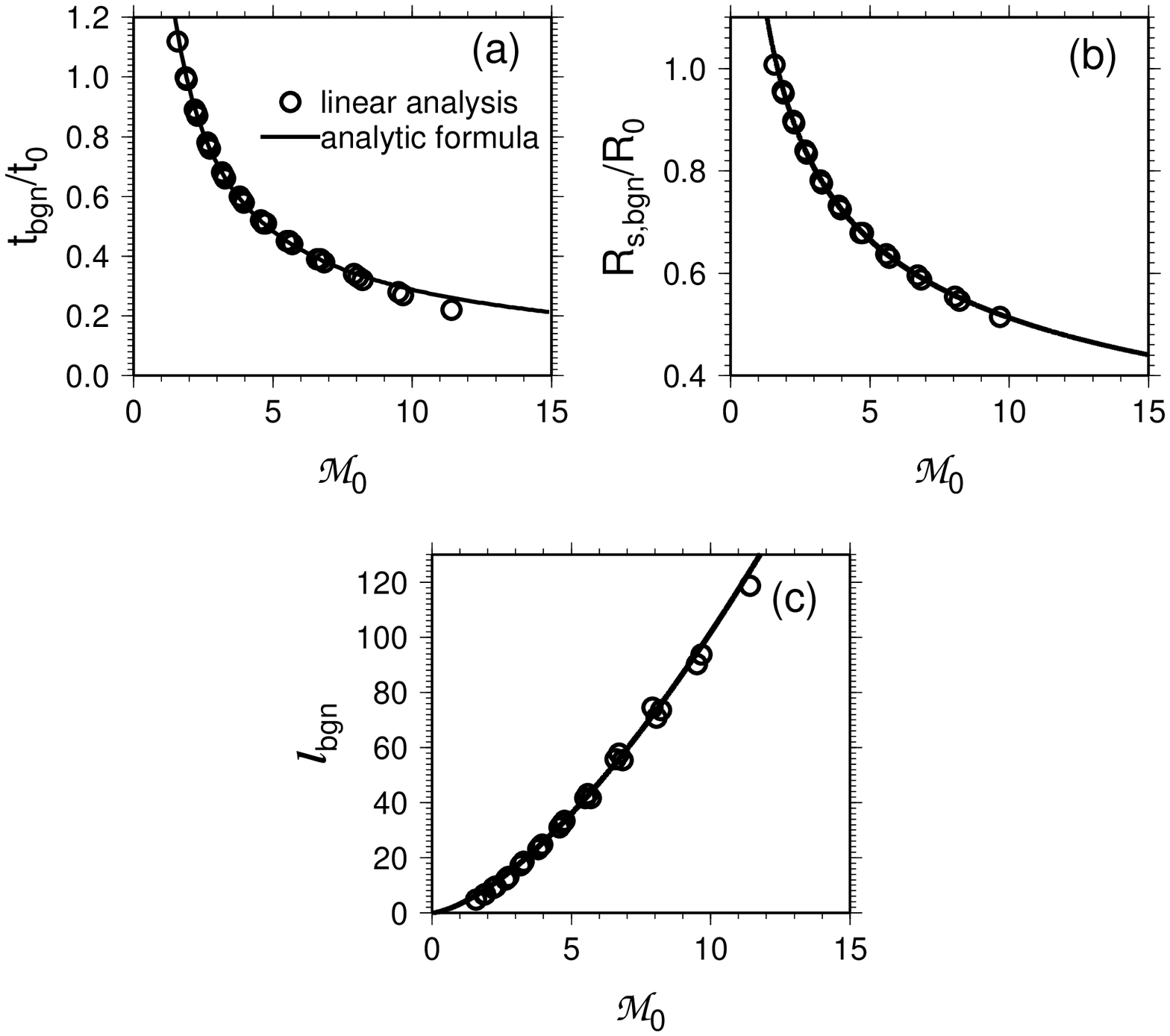}
      \end{center}
\caption{
Dependence of (a)$t_\mathrm{bgn}/t_0$, (b)$R_\mathrm{s,bgn}/R_0$, and 
(c)$l_\mathrm{max}$ on the typical Mach number ${\cal M}_0$.
The open circles correspond to the results of linear analysis. 
The solid lines indicate the analytic formulae.
}
\label{fig:begin cs}
\end{figure}

From Figures \ref{fig:begin}, since the contour lines are almost straight in 
the plane $(\log_{10}n_\mathrm{E},\:\log_{10}Q_\mathrm{UV})$,
quantities are expected to have power-law dependence on $Q_\mathrm{UV}$ and
$n_\mathrm{E}$, or equivalently on ${\cal M}_0$.
In Figure \ref{fig:begin cs},
we plot $t_\mathrm{bgn}/t_0$, $R_\mathrm{s,bgn}/R_0$, and $l_\mathrm{bgn}$
as a function of ${\cal M}_0$ 
for various parameters that correspond to 
the open circles in Figure \ref{fig:begin}. 
The open circles correspond to the results of the linear analysis.
From Figures \ref{fig:begin cs}, one can see that the results of 
the linear analysis are well described by the analytic formulae,
$t_\mathrm{bgn}/t_0=1.62{\cal M}_0^{-0.75}$, 
$R_\mathrm{s,bgn}/R_0=1.2{\cal M}_0^{-0.375}$, 
and $l_\mathrm{bgn}=3.2{\cal M}_0^{1.5}$ that
are plotted by the solid lines. Using Equation (\ref{mach}), the analytic formulae 
are rewritten as 
\begin{equation}
t_\mathrm{bgn} = 0.6\:Q_\mathrm{UV,49}^{-0.11}\:
T_\mathrm{c,10}^{0.375}\:
n_\mathrm{E,3}^{-0.45}\;\mathrm{Myr},
  \label{tbegin}
\end{equation}
\begin{equation}
        R_\mathrm{s,bgn} = 3.4\:
Q_\mathrm{UV,49}^{0.09}\:
T_\mathrm{c,10}^{0.19}\:
n_\mathrm{E,3}^{-0.54}\;\mathrm{pc},
  \label{rbegin}
\end{equation}
and 
\begin{equation}
l_\mathrm{bgn} = 54\;
Q_\mathrm{UV,49}^{0.21}\:
T_\mathrm{c,10}^{-0.75}\:
n_\mathrm{E,3}^{-0.11}.
  \label{lbegin}
\end{equation}
The wavelength $\lambda_\mathrm{bgn}$ and 
the mass $M_\mathrm{bgn}$ that correspond to $l_\mathrm{bgn}$ 
are given by 
\begin{equation}
\lambda_\mathrm{bgn} 
=  0.4\;
Q_\mathrm{UV,49}^{-0.125}\:
T_\mathrm{c,10}^{0.94}\:
n_\mathrm{E,3}^{-0.44}\;\mathrm{pc},
  \label{lambegin}
\end{equation}
and
\begin{equation}
  M_\mathrm{bgn} = 3.5\;
Q_\mathrm{UV,49}^{-0.16}\:
T_\mathrm{c,10}^{2.06}\:
n_\mathrm{E,3}^{-0.42}\;\mathrm{M_\odot},
  \label{mbegin}
\end{equation}

\citet[][W94]{Wetal94} also derived similar formulae by 
using the thin-shell linear analysis similar to E94.
The dependence of Equations (\ref{tbegin})-(\ref{mbegin}) 
on parameters $Q_\mathrm{UV}$, $T_\mathrm{c}$, and $n_\mathrm{E}$ 
shows close agreement with that in W94 
although the numerical factors are different.
Our results show that the GI begins to grow earlier and 
as a result,
the fragment mass is smaller than their results.
In detail, $t_\mathrm{bgn}$, $R_\mathrm{bgn}$, and $\lambda_\mathrm{bgn}$
are roughly half of those in W94, and 
$M_\mathrm{bgn}$ is roughly 1/8 of that in W94.
This is because the stabilization effect of the SF does not work 
on the trailing surface that is the CD as shown in Paper I.
As pointed out in W94, the properties of fragments
are insensitive to $Q_\mathrm{UV}$, and 
mainly depend on $n_\mathrm{E}$ and $T_\mathrm{c}$.
Indeed, in Equations (\ref{lambegin}) and (\ref{mbegin}), 
the dependence of $\lambda_\mathrm{bgn}$ and $M_\mathrm{bgn}$ on $n_\mathrm{E}$ are
close to the Jeans length and the Jeans mass of the shell $(\lambda_\mathrm{J},\;M_\mathrm{J})
\propto n_\mathrm{sh}^{-1/2} \propto n_\mathrm{E}^{-3/7}$.

\subsection{Comparison with Previous Studies}
\citet{E89} investigated the GI in a decelerating shocked layer.
He numerically integrated perturbation equations with respect to time 
by taking into account 
time evolution of the layer and by averaging physical quantities 
over the thickness. 
He has taken into account SF and CD boundary conditions 
on the SF and CD, respectively.
He called the boundary effect on the CD ``pinching force''.
In his linear analysis,
the destabilizing effect of the pinching force does not work efficiently.
One reason is that he used Equation (3) in Paper I as the 
pressure of the HII region although 
the geometry is plane-parallel.
Therefore, the pressure at the CD is underestimated, leading 
to underestimate the pinching force of the CD.
The other reason is the geometry of the gravitational potential.
Asymmetry of density profile of spherical shell is qualitatively different from 
that of plane-parallel layer.
In the plane-parallel geometry, the static symmetric layer peaks at the mid-plane.
If the layer decelerates, the density peak is always closer to the SF than the CD.
Therefore, the tangential flow behind SF erodes the density perturbation more efficiently.
On the other hand, for the case with shells, 
the peaks can be closer to the CD than the SF (see Section \ref{sec:den pro}).
From these reasons, he derived lower growth rates than our results.

\citet{Detal09} simulated the gravitational fragmentation of expanding shells 
confined by temporary constant thermal pressure of hot rarefied gases on both sides.
In their calculation, the motion of the shell is determined so that the pressure at 
the leading surface is the same as that at the trailing surface.
Therefore, the density peak is always around the mid-plane of the shell, and 
the density profile is almost symmetric.
In their calculation, the column density decreases with time because the shell expands keeping the mass fixed.
Therefore, the pressures at the boundaries approach to the peak pressure.
They found that the confining pressure accelerates fragmentation in the later phase, and  
describe this effect as ``pressure-assisted'' gravitational fragmentation. 
This mode is the same as the incompressible mode. 
Recently, taking into account the effect of the thickness of the shell 
approximately, 
\citet{WDPW10} established a semi-analytic linear analysis that
explains results of their SPH simulations.
Different from them, in Paper I, we have taken into account the effect of SF+CD boundary condition 
approximately. 

\subsection{Other Instabilities}
We discuss other potentially important effects that are not analyzed 
in this paper.
\subsubsection{Vishniac Instability}
\citet{V83} found a hydrodynamical overstability on decelerating shells 
confined by the SF on the leading surface and by the CD on the trailing surface 
by using thin-shell linear analysis.
However, the Vishniac instability (VI) did not appear to occur in our simulations. 
This can be explained by the stabilizing effect by expansion.
\citet{VR89} obtained an analytic dispersion relation of the VI of 
the expanding shell without 
self-gravity by taking into account the thickness approximately.
In Section 6 in Paper I, they discussed the VI by applying the analytic 
dispersion relation to the shell driven by the HII region. 
They found that the VI is expected not to be important in the later phase ($t/t_0>0.5$)
since its growth rate is comparable to the expansion rate of the shell radius.
Therefore, the VI was not found in our simulation.
However, they also found that before the beginning time of our simulations ($t/t_0<0.4$), 
the VI is expected to grow rapidly since the Mach number is large 
(see Figure 13 in Paper I). The perturbations quickly reach the nonlinear 
regime and saturate as a subsonic transverse flow \citep{MN93}. 
As a results, a small scale subsonic turbulence whose angular wavelength of 
$l=10^2\sim10^3$ is expected to be induced. The consequence of VI may correspond to the 
increase of the initial perturbations in our 3D simulation.

\subsubsection{Thermal Instability}\label{sec:TI}
In this paper, we assume that the fluid evolves keeping its temperature fixed.
However, in real interstellar medium (ISM), shock-heated gas cools via radiative cooling.
It is well known that the cooling ISM is often thermally unstable.
In such a case, during the cooling condensation, the layer is expected to 
fragment into dense clouds with various scales \citep{IT09}.
\citet{KI02} investigated the propagation of a shock wave into the warm 
neutral medium. They found that cold cloudlets move with supersonic 
velocity dispersion in surrounding 
warm gas in the postshock region. 
The velocity dispersion is larger than the sound speed of cloudlets while 
it is smaller than the sound speed of surrounding warm gas.
Since the shocked molecular cloud is also thermally unstable, the thermal instability 
drives supersonic turbulence inside the shell, and 
cold molecular cloudlets move with the supersonic velocity dispersion.
This situation is quite different from the isothermal gas that has been considered 
in this paper.
The supersonic turbulence is expected to be important in contrast to 
the subsonic turbulence
driven by the VI, and considerably modify the evolution.
The mass of the cold molecular cloudlets increases with time through the accretion of 
surrounding gas and the coalescence between cloudlets.
The supersonic turbulence may provide effective turbulent pressure, 
while coalescence of cold cloudlets may decrease turbulent pressure. 
We will address the effect of the thermal instability on the 
GI of the expanding shells in our next work.

\subsection{Ionization Front}
In the paper, we did not solve the radiative transfer equation of the ionizing photons.
Instead, we assumed that the trailing surface is the CD instead of the 
ionization front (IF).
There are two effects of the ionization on the GI.
One is that a fraction of the shell becomes ionized gas as HII region expands.
Therefore, our simulations overestimate the column density of the shell.
The swept-up mass by the SF is given by 
\begin{equation}
        M_\mathrm{sw} = \frac{4\pi}{3}\rho_\mathrm{E} R_\mathrm{s}^3.
        \label{swept mass}
\end{equation}
On the other hand, the ionized mass is given by 
\begin{equation}
        M_\mathrm{ion} = \frac{4\pi}{3}\rho_\mathrm{i} R_\mathrm{s}^3,
        \label{ion mass}
\end{equation}
where $\rho_\mathrm{i}$ is the density of the ionized gas, and 
we neglect the difference between radii of the IF and SF.
The real shell mass is given by $M_\mathrm{sw}-M_\mathrm{ion}$.
Inside the HII region, the balance between the ionization and the recombination
is approximately established. Therefore, the density of the HII region becomes
\begin{equation}
    \rho_i  
    = \left( \frac{3m^2 Q_\mathrm{UV}}{4\pi \alpha_\mathrm{B} R_\mathrm{s}^3} \right)^{1/2}
    = \rho_\mathrm{E}\left( \frac{R_\mathrm{ST}}{R_\mathrm{s}} \right)^{3/2},
    \label{ion den}
\end{equation}
where $m$ is the mean weight of the gas particles and we use Equation (\ref{stremgren}).
Therefore, from Equations (\ref{swept mass})-(\ref{ion den}), 
the ratio of $M_\mathrm{ion}$ to $M_\mathrm{sw}$ is 
\begin{equation}
        {\cal R}\equiv \frac{M_\mathrm{ion}}{M_\mathrm{sw}}  
    = \left( \frac{R_\mathrm{ST}}{R_\mathrm{s}} \right)^{3/2}.
\end{equation}
If ${\cal R}\ll 1$, the effect of the ionization is negligible.
In Figure 1 in Paper I (also see Figure \ref{fig:surpos}), the Str{\"o}mgren radius is 
as large as $0.1R_0$ in various parameters $(Q_\mathrm{UV},\;n_\mathrm{E})$.
At the beginning time of our simulations, $t=0.4t_0$ $(R_\mathrm{s}\sim0.6R_0)$, ${\cal R}\sim 
(0.1/0.6)^{3/2}=0.07$. As the shell expands, ${\cal R}$ decreases. 
At $t=t_0$, ${\cal R}\sim (0.1/10)^{3/2}\sim 0.03$.
Therefore, the contribution of the ionization is limited to only several percent.

The other is that the IF induces an instability on the shell.
The D-type IF is analogous to a combustion front that 
is unstable to corrugational deformations \citep{LL87}.
\citet{V62} found that the D-type IF is unstable. 
In this instability, the most unstable scale is comparable to the thickness 
of the IF.
Moreover, \citet{GF95} showed that the VI is strongly modified by the IF.
The shell rapidly fragments and finger-like 
structures are generated in their two-dimensional 
simulations.
Note, however, that they assumed the power-law density distribution $\propto r^{-w}$ of 
the ambient gas with the uniform density core in the center with $0.2$ pc, 
where parameter $w$ is set to 2.
The expansion of the IF depends on the power-law index $w$ in the 1D model.
If $w>1.5$, even if the SF is emerged in front of the IF in the uniform density core, the 
IF quickly gets ahead of the SF, and eventually 
the whole of the shell and the ambient gas are ionized.
Moreover, \citet{H07} investigated the expansion of the IF and the dissociation front,
and showed that the star formation is expected to be suppressed when $w>1$.
This is because the column density decreases as the shell expands and the 
FUV photon easily escapes in front of the shock.
Therefore, model by \citet{GF95} corresponds to the case where 
the triggered star formation is not simply expected.
We need to investigate the expansion of the IF for the case with the shallower density 
profile $w<1$ in detail 
using the radiative multi-dimensional simulations taking into account the self-gravity.
Moreover, \citet{W99} discovered shadowing 
instability in the R-type IF before emerging of the SF.
This instability may also disturb the shell in the very early phase of evolution.

\subsection{Magnetic Field}
It is well known that the magnetic field is important in the dynamics of the ISM although 
it is neglected in this paper.
\citet{NIM98} investigated the GI of a pressure-confined 
isothermal layer threaded by 
uniform magnetic field by using linear analysis.
They found that the magnetic field cannot stabilize the GI.
Moreover, they found the layer fragment in filamentary clouds whose 
longitudinal axis 
is parallel or perpendicular to the magnetic field 
depending on the thickness of the layer.
However, their analysis is restricted to the static magnetized layer.
It is important to investigate stability of magnetized shocked layer or shell
in dynamical modeling.

\section{Summary}\label{sec:summary}
In this paper, we have investigated the GI of the expanding shell using 
3D numerical simulation with high resolution.
We summarize our results as follows:

\begin{enumerate}
    \item The GI begins to grow earlier than the 
          prediction from the linear analysis
          based on the thin-shell approximation \citep{E94, Wetal94}.
          During the development of the GI, 
          the contact discontinuity highly deforms while the shock front remains almost flat.
          These all features are consistent with the prediction 
          from \citet{IIT11}.

   \item The GI is expected to begin to grow at an epoch 
         when the growth rate of the GI becomes larger than 
         the evolutionary rate owing to the expansion and the mass accretion (see Equation (\ref{frag criterion}))
         Using the results of the linear analysis, 
         we have derived useful approximate analytic formulae (Equations (\ref{tbegin})-(\ref{mbegin}))
         for the fragment scale
         and the epoch when the GI starts.
\end{enumerate}

This paper and Paper I revisit the fragmentation of the expanding shells by using both
the 3D simulation and the linear analysis.
Since the gas is subject to heating owing to far-UV photon, the gas temperature is 
larger than 10 K \citep{HI05,HI06}.
When $T_\mathrm{c}=30$ K ($50$ K), the GI with mass scale of 
$M_\mathrm{bgn}\sim 34\:M_\odot\:$ ($96$ $M_\odot$)  
begins to grow at $t_\mathrm{bgn}\sim0.9$ Myr (1.1 Myr) 
($Q_\mathrm{UV}=10^{49}$ s$^{-1}$, $n_\mathrm{E}=10^3$ 
cm$^{-3}$) from Equations (\ref{tbegin}) and (\ref{mbegin}).
The fragment mass is expected to be larger than $M_\mathrm{bgn}$ because 
it increases through the gas accretion.
Moreover, perturbations with larger scale than the most unstable mode
is also unstable. Therefore, the larger scale perturbations gradually grows, and 
small scale fragments may coalesce into larger one, and 
the total number of fragments may decrease.
The predicted masses are roughly comparable 
to (or are slightly smaller than) observational 
masses of dense cores (dozens $\sim$ hundreds M$_\odot$) \citep[e.g.,][]{Detal03,Zetal06} 
although our model is based on very idealized situation, such as uniform ambient 
gas and isothermal equation of state.

\acknowledgments
We thank the referee for very careful reading and 
many constructive comments that improved this paper significantly.
Numerical computations were carried out on PC cluster at Osaka University Cybermedia Center,
and Cray XT4 at the CfCA of National Astronomical Observatory of Japan.
This work was supported by Grants-in-Aid for Scientific Research
from the MEXT of Japan (K.I.:22864006; S.I.:18540238 and 16077202), 
and Research Fellowship from JSPS (K.I.: 21-1979). 
The page charge of this paper is supported by CfCA


\appendix
\section{Effect of Discontinuity in Gravitational Force}\label{app:grav}
\begin{figure}[htpb]
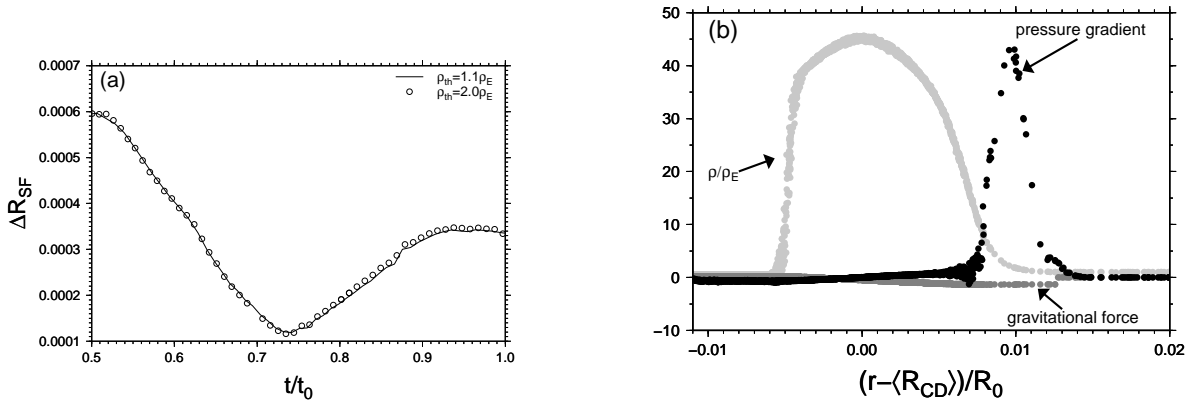

      \begin{center}
     \begin{tabular}{cc}
         \includegraphics[width=7.5cm]{f14a.eps}
         &
         \includegraphics[width=7.5cm]{f14b.eps}
     \end{tabular}
      \end{center}
\caption{
a) Time evolution of the displacement of the SF. 
The solid line and the open circles correspond to $\rho_\mathrm{th}=1.1\rho_\mathrm{E}$ and 
$2\rho_\mathrm{E}$, respectively.
b) Snapshots of the pressure gradient (the light gray circles), and the 
gravitational force (the dark gray circles) along the $r$-direction in the simulation unit.
The density distribution is superimposed by the filled circles.
The abscissae are the distance from the CD.
}
\label{fig:acc grav}
\end{figure}

In this paper, the gravitational force is neglected outside $R_\mathrm{grav}$ (see Section \ref{sec:ini bou}).
We investigate how the sudden change in the gravitational force at $R_\mathrm{grav}$ influences our results.
To see this, 
we focus on the time evolution of the displacement of the SF $\Delta R_\mathrm{SF}$ whose definition 
is the same as $\Delta R_\mathrm{CD}$ (see Section \ref{sec:focus}).
When $\Delta R_\mathrm{SF}$ is evaluated, 
the position of the SF along ${\bf n}_i$ is determined as the density coincides with $1.1\rho_\mathrm{E}$.
Figure \ref{fig:acc grav}(a) shows that the time evolution of $\Delta R_\mathrm{SF}$ for $l=52$ in the 
HM model. The solid line and the open circles correspond to $\rho_\mathrm{th}=1.1\rho_\mathrm{E}$ and 
$2\rho_\mathrm{E}$, respectively. One can see that they agree very well.
Therefore, our results are insensitive to the position of the discontinuity of the gravitational force.
This is because the pressure gradient is much larger than the gravitational force at the shock transition 
layer. This is clearly seen in 
Figure \ref{fig:acc grav}(b) that shows the pressure gradient, the gravitational force and the density at
$t=t_0$ for $l=52$ in the simulation unit. The SPH particles around $\phi=0$ is plotted.
The threshold density is set to $1.1\rho_\mathrm{E}$.
One can see that the gravitational force has the jump around $(r-\langle R_\mathrm{C}\rangle)/R_0\sim0.013$ 
that corresponds to $R_\mathrm{grav}$ in Figure \ref{fig:acc grav}(b).
From Figure \ref{fig:acc grav}(b), the discontinuity in the gravitational force is much smaller than the pressure 
gradient because the pressure suddenly changes at the shock transition layer.  

Therefore, it is concluded that 
the existence of the discontinuity in the gravitational force is not important 
as long as the discontinuity is inside the shock transition layer.

\end{document}